\documentstyle[pre,aps]{revtex}
\input psfig
\def\fg{\epsilon_d}
\def\gradv{\nabla}
\def\KK{{\overline \kappa}}
\def\tv{{\bf t}}
\def\mub{{\overline \mu}}
\def\fd{{\epsilon_d}}
\def\lv{{\bf l}}
\def\Xv{{\bf X}}
\def\Rv{{\bf R}}
\def\uv{{\bf u}}
\def\Ha{{\cal H}}
\def\lvp{{\lv^{\prime}}}
\def\av{{\bf a}}
\def\xv{{\bf x}}
\def\bv{{\bf b}}
\def\nv{{\bf n}}
\def\qv{{\bf q}}
\begin{document}
\twocolumn[\hsize\textwidth\columnwidth\hsize\csname@twocolumnfalse%
\endcsname
\title{Chiral Discotic Columnar Phases in Liquid Crystals}
\author{Gu Yan and T.C. Lubensky}
\address{Department of Physics and Astronomy, University of Pennsylvania,
Philadelphia, PA 19104}
\draft
\date{\today}
\maketitle
\begin{abstract}
We introduce a model to describe columnar phases of chiral discotic liquid
phases in which the normals to disc-like molecules are constrained to lie
parallel to columnar axes. The model includes separate
chiral interactions favoring, respectively,
relative twist of chiral molecules along the axes of the columns and
twist of the two-dimensional columnar lattice.  It also includes a coupling
between the lattice and the orientation of the discotic molecules.
We discuss the instability of the aligned hexagonal lattice phase to the
formation of a soliton lattice in which molecules twist within their
columns without affecting the lattice and to the formation of a moir\'{e}
phase consisting of a periodic array of twist grain boundaries
perpendicular to the columns.
\end{abstract}
]
\section{Introduction}
\par
Chirality gives rise to a rich variety of modulated equilibrium phases in
liquid crystals\cite{prost},
including the cholesteric, smectic-$C^*$, TGB, and blue
phases.  To date most theoretical and experimental work has focussed on
systems composed of molecules that adopt rod-like conformations that favor
the formation of uniaxial smectic phases. Discotic chiral molecules favoring
the formation of columnar phase have, however, been
synthesized\cite{malthete,bock,scherowsky,swager}.  They produce discotic
cholesteric phases\cite{malthete} and columnar phases with chiral structure.
They also exhibit interesting ferroelectric properties\cite{bock,scherowsky}.
\par
\begin{figure}
\centerline{\psfig{figure=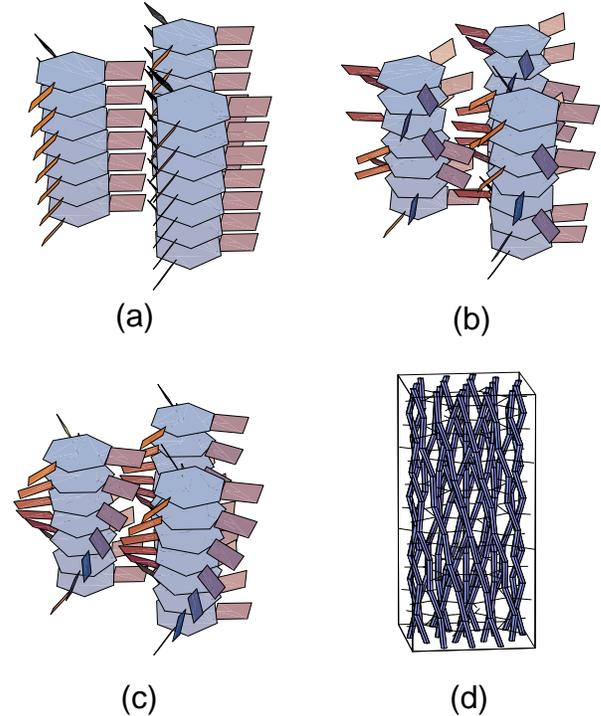}}
\caption{Some chiral discotic columnar phases: (a) A uniform
orientationally ordered phase in which there is not twist.  This twist has
been expelled as in the smectic-$A$ phase of chiral molecules, (b) An
orientationally disordered or ``plastic" columnar phase.  The is no
long-range orientational order of the molecules, but the hexagonal
columnar structure remains intact. (c) A cholesteric-like phase in which
molecules rotate in a helical fashion within each column while maintaining
phase coherence between columns. (d) The moir\'{e} phase in which the
columns themselves rotate about the $z$ axis.  Rotations occur in discrete
jumps about twist grain boundaries. (From Ref. \protect\onlinecite{KamNel})}
\label{fig1}
\end{figure}

Chirality favors twisted structures.  In chiral smectic phases, twist can
be expelled altogether in the smectic-$A$ phase, it can appear as molecular
twist in the smectic-$C^*$ phase, or it can appear as layer twist in the TGB
phases\cite{RennLub}. Columnar phases can exhibit the analog
of all of these phases and some
phases with no analog in smectic systems. Some possible columnar discotic
phases are shown in Fig.\ \ref{fig1}.
If chiral forces are
sufficiently weak, the lattice structure of the columnar phase can simply
expel twist.  If the coupling between molecular chirality and the
columnar lattice is sufficiently strong\cite{KamNel}, chirality can
in principle induce a tilt-grain-boundary phase,
analogous to the TGB phase in smectics,
with rotation axis perpendicular to the columns, or
a moir\'{e} phase with
rotation axis parallel to the columns.  In the former phase, there is a
periodic lattice of grain boundaries separating rotated regions of aligned
columns.  In the latter, there is a periodic lattice of grain boundaries
perpendicular to the columns across which the hexagonal columnar lattice
undergoes discrete rotations. In columnar phases, molecules can twist
within their columns without destroying the lattice structure.  Such
behavior is possible in columnar systems because each molecule sits in a
fairly symmetric environment, and it can rotate without greatly disturbing
its neighbors.  In smectic phases, the rotation of a molecule within a
layer would lead to enormous disruptions and would be energetically very
costly.  Phases with spontaneous chiral symmetry breaking in
which propeller-like
molecules rotate in different directions in different columns have been
observed experimentally\cite{heiney} and modeled theoretically\cite{caille}. In
chiral systems, the direction of preferred rotation in all columns is set
by the underlying molecular chirality. The pattern of molecular rotation
in these phases can be described as a soliton lattice, and we will refer
to them as soliton phases.
\par
In this paper, we will introduce a fairly general model for chiral
columnar phase that has soliton and moir\'{e} phases in addition to
phases with expelled twist.  To keep the model as simple as possible,
we restrict the normal to the disc-like molecules to lie parallel to the
columnar axes.  We will, therefore, not be able to discuss the
tilt-grain-boundary phase or ferroelectric chiral discotics.  In a future
publication, we will study a variety of phases that can result with this
constraint is relaxed. Our model begins with interacting chiral molecules in a
discotic columnar phase.  Its Hamiltonian has an elastic part associated with
distortions of the lattice.  It has terms favoring parallel alignment of
neighboring molecules and chiral terms favoring both molecular and lattice
rotation.   It also has crystal-field terms coupling molecular orientation to
the columnar lattice and defining a twist penetration depth
$\lambda_{\theta}$ analogous to that of smectic systems.   If lattice
distortions are prohibited, our model is essentially identical to that studied
by the Sherbrooke group\cite{caille}.   We assume that the low-temperature,
weak chirality, ground state of our Hamiltonian is the ordered
``ferromagnetically" aligned state shown in Fig.\ \ref{fig1}a, though other
large unit-cell states are possible\cite{caille}.
As chirality is increased, the aligned state
can become unstable either to a soliton lattice if the crystal-field
coupling is weak or to a moir\'{e} phase if it is strong.  Our primary aim
will be to determine the critical chirality for these two instabilities.
Our model, however, has the potential for more complicated phases and very
complex phase diagrams.  For example, a mixed moir\'{e}-soliton phase in
which there is molecular twist relative to the lattice between grain
boundaries could exist.
Or the soliton phase could melt altogether to a ``plastic"
discotic phase with no orientational long-range order (Fig.\
\ref{fig1}b).  This phase could them become unstable with respect to the
formation of a moir\'{e} phase.
\par
Though our model is motivated by chiral discotic liquid crystals, it can,
with proper interpretation, be applied to aligned chiral polymers such as
DNA\cite{DNA}.  Aligned polymers can form a hexagonal lattice perpendicular to
the direction of alignment.  A polymer like DNA is a tightly wound double
helix.  This structure makes it unlikely for it to form an orientationally
ordered phase analogous to the discotic ground state shown in Fig.\
\ref{fig1}a.  Rather the phases of the molecular helices on
neighboring molecules will be random: there will be no long-range order in
molecular orientations perpendicular to the direction of polymeric
alignment. The ground state will be equivalent to the disordered states
shown in Fig.\ \ref{fig1}b, though each molecule will have an average
twist.  Thus, there is no analog of the soliton phase in DNA, and the
moir\'{e} phase is driven by chiral terms favoring lattice rotation rather
than molecular rotation.
\par
The outline of this paper is as follows.  In Sec. II, we define the
model and discuss its continuum limt.  In Sec. III we discuss
instabilities toward the moir\'{e} phase.  This analysis differs from that
of reference \onlinecite{KamNel} because of the finite twist penetration
depth.  Our results are, however, almost identical to those of that
reference.  In Sec. IV, we discuss the soliton phase and arrive at a
criterion determining whether the soliton or the moir\'{e} phase will form.
\section{The Model}
\subsection{The Lattice Model}
\par
\begin{figure}
\centerline{\psfig{figure=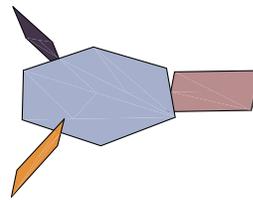}}
\caption{Schematic representation of a chiral discotic molecules with
$C_3$ symmetry.  The three vanes coming off the hexagonal core are all
tilted in the same direction relative to the normal like the blades of an
airplane propeller.}
\label{chiralmolfig}
\end{figure}

Columnar forming chiral molecules can come in many forms.  We will limit
our discussion to molecules such as that shown in Fig.\ \ref{chiralmolfig}
with $C_3$ symmetry.  This molecule is similar to some that have
recently been synthesized\cite{swager} and to those spontaneously formed
in the experiments of Heiney\cite{heiney}.  Since we are interested
principally in the competition between soliton and moir\'{e} phases, we will
assume that the molecular normal always aligns along the columnar
direction.  We will, therefore, not be able to discuss the formation of
the tilt-grain-boundary phase.
We use a mixed continuum-lattice description of the
columnar phase.  The discotic columns form a two-dimensional
hexagonal lattice with lattice parameter $a$. They are labeled by an
index $\lv$. Distance
parallel to the columns is specified by the continuous coordinate $z$.
The orientation of molecules at position $z$ in column $\lv$ is
specified by the angle $\theta_{\lv} ( z)$. The column coordinates are given
by $\Xv_{\lv} ( z ) = \Rv_{\lv} + \uv(\lv, z )$ , where $\Rv_{\lv}$
is the equilibrium two-dimensional lattice coordinate and $\uv(\lv,z)$
is displacement from equilibrium that can depend on both $z$ and
$\lv$.
\par
The Hamiltonian for our lattice model can be divided into an elastic part
${\overline \Ha}_{\rm el}$, an angle part ${\overline\Ha}_{\theta}$,
and a chiral part
${\overline \Ha}^*$.  The elastic Hamiltonian is the standard one for a
columnar structure.  In the harmonic limit, it is
\begin{eqnarray}
{\overline \Ha}_{\rm el}
&= &{1 \over 2} \int dz\sum_{\lv,\lvp} K_{ij} ( \lv , \lvp )
u_i ( \lv , z ) u_j(\lvp,z) \nonumber \\
& & + {1 \over 2} \int dz \sum_{\lv} \KK
[\partial_z^2 \uv( \lv ,z )]^2 .
\label{Hel}
\end{eqnarray}
The first term in this expression, with $K_{ij} ( \lv , \lvp )$
an elastic constant tensor, is the familiar elastic energy of a
two-dimensional harmonic lattice of columns.  The second term,
with ${\overline \kappa}$ a bending rigidity, measures
the energy of bending the columns.  We assume that neighboring
molecules want to be parallel in the absence of chirality.  Furthermore, there
are couplings between lattice distortions and molecular rotation and a
preferred orientation of the molecules relative to the lattice.  All of these
effects are incorporated into the model Hamiltonian
\begin{eqnarray}
{\overline \Ha}_{\theta}& =& - {1 \over 2} A \int dz \sum_{\lv, \lvp} \cos
[3\theta_{\lv} ( z ) - 3 \theta_{\lvp} ( z )] \nonumber \\
& & - {1 \over 2} B \int dz
\sum_{\lv,\lvp} \cos [6 \theta_{\lv} ( z ) - 6 \phi_{\lv,\lvp}(z)] \nonumber
\\
& & - {1 \over 2} C \int dz \sum_{\lv,\lvp} \cos [3 \theta_{\lv} ( z ) + 3
\theta_{\lvp } ( z) - 6 \phi_{\lv, \lvp} ( z)] , \nonumber \\
& & + {1 \over 2} \KK_{\theta}
\sum_{\lv}\int dz [\partial_z \theta_{\lv} ( z ) ]^2
\label{Htheta}
\end{eqnarray}
where the sum $\lv,\lvp$ is over nearest neighbor columns in the lattice
and
\begin{equation}
\phi_{\lv,\lvp} ( z ) = \tan^{-1}\left( {a_y^{\lv,\lvp} + u_y (\lv , z ) -
u_y(\lvp , z) \over a_x^{\lv,\lvp} +
u_x (\lv ,z ) - u_x ( \lvp,z)} \right)
\label{philat}
\end{equation}
is the angle the bond between column $\lv$ and
$\lvp$ makes with the $x$-axis at $z$ and $\av^{\lv,\lvp}$ is the
equilibrium lattice vector connecting those columns.
The Hamiltonian ${\overline
\Ha}_{\theta}$ is invariant under $\theta_{\lv} \rightarrow \theta_{\lv} +
(2 \pi n/3)$ for any interger $n$, as required by the $C_3$ molecular
symmetry.  It is also invariant under rotations of the lattice by $2 \pi/6$ and
under simultaneous rotations of the molecules and the lattice through
arbitrary angles.
When lattice distortions are prohibited, ${\overline
\Ha}_{\theta}$ is very similar to that studied by the Sherbrooke
group\cite{caille}.
\par
Finally chiral interactions along a given column favor molecular rotation,
and chiral interactions between molecules in different columns favor
lattice rotation.  We introduce two chiral terms in our model to describe
these effects:
\begin{eqnarray}
{\overline\Ha}_{\theta}^* & = & -{\overline \gamma}_{\theta}
\int dz \sum_{\lv} \partial_z \theta_{\lv} ( z)
\nonumber \\
{\overline\Ha}_{\phi}^* & = & - {\overline \gamma}_{\phi}
\sum_{\lv,\lvp}\int dz \partial_z \phi_{\lv,
\lvp} .
\end{eqnarray}
In the ordered phase of discotic systems, the dominant twist comes from
${\overline\Ha}_{\theta}^*$, and we may assume
$\gamma_{\theta} \gg \gamma_{\phi}$.
In polymeric systems or in the orientationally disordered phase,
$\gamma_{\theta}$ is zero, and any rotation is
induced by ${\overline\Ha}_{\phi}^*$
\par
\subsection{The Continuum Limit}
\par
When spatial variations are slow on the scale of the lattice spacing, we may
expand the lattice Hamiltonian in gradients of lattice displacements $\uv$
and angles $\theta$ and replace the lattice sum by a continuum integral.
To this end, we replace
$\uv ( \lv ,z )$ by $\uv(\xv)$ and $\theta_{\lv} ( z )$ by
$\theta( \xv )$, where $\xv = (\xv_{\perp},z)$
with $\xv_{\perp}$ the coordinate perpendicular to $z$, and we set
$\int dz \sum_{\lv} \rightarrow a^{-2} \int d^3x$.
In this limit, the lattice angle $\phi_{\lv,\lvp} ( z)$ can be expanded
about its equilibrium value of $\phi_{\lv,\lvp}^0 ( z)$ as
\begin{eqnarray}
\delta \phi_{\lv,\lvp}(z) & = & \phi_{\lv,\lvp} ( z ) - \phi_{\lv,\lvp}^0 (z)
\nonumber
\\
&= &
a^{-2} [ a_x^{\lv,\lvp} a_i^{\lv,\lvp} \partial_i u_y ( \xv )
-a_y^{\lv,\lvp} a_i^{\lv,\lvp} \partial_i u_x ( \xv )] .
\end{eqnarray}
The Hamiltonian ${\overline \Ha}_{\theta}$ does not depend on
the equilibrium angle $\phi_{\lv,\lvp}^0$ because the latter
is an integral multiple of $2 \pi /6$. The average over nearest neighbor
lattice sites of $\delta \phi_{\lv, \lvp} (z )$ is the hexatic angle
$\phi_6 ( \xv )$:
\begin{equation}
\phi_6 ( \xv ) = \case{1}{2} \epsilon_{ij}\partial_i u_j = {1 \over 6}
\sum_{\lvp} \delta \phi_{\lv,\lvp} ( z ) ,
\end{equation}
where $\epsilon_{ij}$ is the anti-symmetric two-dimensional tensor with
$i$ and $j$ running over $x$ and $y$.
In addition,
\begin{equation}
\sum_{\lvp} [\delta \phi_{\lv\lvp}^2 ( z ) ] = 6 \phi_6^2 + 3 u_{ij} u_{ij} -
\case{3}{4} u_{ii}^2 ,
\end{equation}
where $u_{ij} = (\partial_i u_j + \partial_j u_i )/2$ is the symmetrized
strain.
With this information, we can express the continuum-limit Hamiltonian as
a sum of terms:
$\Ha = \Ha_{\rm el} + \Ha_{\theta} + \Ha_g + \Ha^*$.
The elastic Hamiltonian
\begin{equation}
\Ha_{\rm el} = \case{1}{2}\int d^3 x [\lambda u_{ii}^2 + 2 \mu u_{ij}u_{ij} +
K (\partial_z^2 \uv )^2] ,
\end{equation}
is the standard continuum elastic Hamiltonian for a hexagonal
columnar system.  Here
$K = \kappa /v_0$, where $v_0= \sqrt{3} a^2/4$ is
the area of an hexagonal unit cell, and the Lam\'{e} coefficients
$\lambda$ and $\mu$ are determined by the continuum limit of $K_{ij} ( \lv
, \lvp )$ and by the the parameters $B$ and $C$ in
${\overline\Ha}_{\theta}$ through the second order expansion in
$\delta\phi_{\lv,\lvp}$.   The angle Hamiltonian is simply that of an
anisotropic $xy$ model:
\begin{equation}
\Ha_{\theta} = \case{1}{2}\int d^3 x [K_{\theta}^{\perp}
(\gradv_{\perp}\theta )^2 + K_{\theta}^{||} (\partial_z\theta)^2 ]  ,
\end{equation}
where $K_{\theta}^{||} = \KK_{\theta}/v_0$, $K_{\theta}^{\perp} =
9(A+C)/2$, and $\gradv_{\perp} = (\partial_x, \partial_y, 0)$.
The coupling term $\Ha_g$ is in the small $\theta - \phi_6$
limit is
\begin{equation}
\Ha_g = {1 \over 2} g \int d^3 x (\theta - \phi_6 )^2
\end{equation}
where $g = 108 (B + C )/a^2$ and where the final form is valid in the
small $\theta$ limit.
Finally, the chiral energy becomes
\begin{equation}
\Ha^* = - \gamma_{\theta} \int d^3 x \partial_z \theta -
\gamma_{\phi} \int d^3 x \partial_z \phi_6 ,
\end{equation}
where $\gamma_{\theta} = {\overline \gamma}_{\theta}/v_0$
and $\gamma_{\phi} = {\overline \gamma}_{\phi}/v_0$.
\par
The continuum Hamiltonian has a
structure imposed by rotational invariance: the coupling term
$\case{1}{2}g (\theta - \case{1}{2} \epsilon_{ij} \partial_i u_j )^2$ is
invariant under simultaneous rotations of the lattice and molecular
orientations.  It is the analog for columnar systems of the invariant
coupling\cite{smcoup} $(\delta \nv + \gradv u )^2$ of smectic-$A$ liquid
crystals, where $\delta \nv$ is the deviation of the
Frank director form its
equilibrium orientation. The columnar phase, like the smectic-$A$ phase
tends to expel molecular twist.  In the ground state, molecules align along
preferred crystal axes with $\theta = 0$.  A twist in $\theta$ relative to
the lattice at $z = 0$ will decay to zero in  twist penetration depths
\begin{equation}
\lambda_{\theta}^{||} = \sqrt{K_{\theta}^{||}/ g} , \qquad
\lambda_{\theta}^{\perp} = \sqrt{K_{\theta}^{\perp} /g} ,
\label{lambdag}
\end{equation}
which tend to zero in this strong coupling, $g \rightarrow \infty$
limit.  Another important length is
\begin{equation}
\lambda= \sqrt{K/ \mu}
\end{equation}
giving the length scale over which bend deformations heal.
\par
Two limits of the continuum Hamiltonian deserve special attention.  The
first is that in which there is no long-range orientational order.
In this limit, $\theta$ looses its meaning, and only terms involving $\uv$
remain. The Hamiltonian becomes $\Ha_{\rm el} + \Ha_{\phi}^*$ with, in
particular, $g = 0$.   This is the limit studied in
\onlinecite{KamNel}.  The second is the strong-coupling $g
\rightarrow \infty$ limit in which $\theta$ and $\phi_6$ are forced to be
equal.  This leads to $\Ha^* = - (\gamma_{\theta} + \gamma_{\phi} ) \int d^3 x
\partial_z \phi_6$ and $\Ha_{\theta} = \case{1}{8}\int d^3 x [
K_{\theta}^{\perp} (\gradv_{\perp} \epsilon_{ij}\partial_i u_j )^2 +
K_{\theta}^{||} (\partial_z\epsilon_{ij}\partial_i u_j )^2 ]$.
Thus, except for a $(\partial^2
u)^2 $ correction, the form of the Hamiltonian with $g = \infty$ is identical
to that for $g = 0$. We should expect, therefore, that the critical values of
chiral couplings leading to the moir\'{e} phase will have similar forms
but differ in magnitude in the two limits and that there will be a smooth,
non-singular interpolation between these limits as a function of $g$.

\section{The Soliton Phase}
\par
If the coupling $g$ between molecular and lattice rotations is weak,
molecules will be able to rotate relative to the fixed lattice.
We can describe this situation by a Hamiltonian depending only on $\theta$
and not on $\uv$: $\Ha = \Ha_{\theta} + \Ha_g + \Ha_{\theta}^*$, where
\begin{equation}
\Ha_g = - \case{1}{36} g \int d^3 x \cos 6 \theta .
\end{equation}
This Hamiltonian is a chiral Sine-Gordon model that is equivalent to
the continuum limit of the Frenkel-Kontorowa model\cite{FK} and that describes
a cholesteric liquid crystal in an external field.
The chiral term $\Ha_{\theta}^*$ favors twist that the angle elastic
term $\Ha_{\theta}$ oposes.  Twist sets in when the coupling $g$ exceeds
the critical value necessary to nucleate a single soliton, which produces
a rotation of $\theta$ through an angle of $\pi/3$ from one end of the
sample to another. The energy per
unit area of a single soliton is
\begin{equation}
\sigma = {2 \over 9} \sqrt{K_{\theta}^{||} g} = {2 \over 9} g
\lambda_{\theta}^{||} .
\end{equation}
The chiral energy gain arising from a single soliton is $\Ha_{\theta}^* =
- A (\pi/3) \gamma_{\theta}$, where $A$ is the area. Thus
the total energy per unit area of a single soliton is
\begin{equation}
f_s = [\sigma - (\pi \gamma_{\theta}/3) ].
\end{equation}
The critical chiral coupling constant is, therefore,
\begin{equation}
\gamma_{\theta}^s = {2 \over 3 \pi} \sqrt{g K_{\theta}^{||}} .
\end{equation}
For $\gamma_{\theta} > \gamma_{\theta}^{s}$, there will be a soliton
lattice in $\theta$ with a lattice spacing that decreases with increasing
$\gamma_{\theta}$.  This is a helical state with a pitch equal to the
soliton lattice spacing.
\section{The Moir\'{e} Phase}
\par
\begin{figure}
\centerline{\psfig{figure=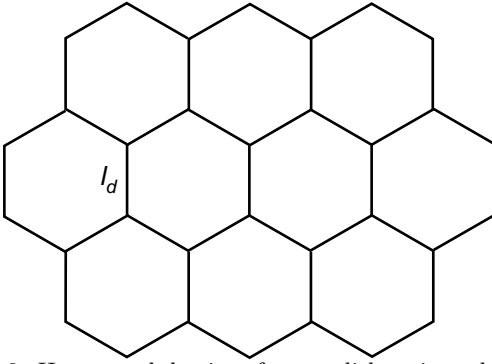}}
\caption{Honeycomb lattice of screw dislocations that comprise twist grain
boundaries in the moir\'{e} phase}
\label{honeycomb}
\end{figure}

The moir\'{e} phase consists of a periodic array of twist grain
boundaries perpendicular to the columnar axis across which the orientation
of the hexagonal lattice rotates in discrete jumps.  The twist grain
boundary is a honeycomb lattice of screw dislocations. This phase forms
when the energy cost of creating a low-angle grain boundary is just
counterbalanced by the twist energy gain arising from $\Ha^*$.
Let $l_b$ be the distance between grain boundaries and $l_d$ be the
length of the side of the hexagonal unit cell of the honeycomb
dislocation lattice (see Fig. \ref{honeycomb}).  The total length of
screw dislocations in a lattice of $N$ cells is $L=3Nl_d$.  The total area of
this lattice is $A=3Nl_d^2/2 \sqrt{3}$.  In the limit of large lattice
spacing (i.e., large $l_b$ and $l_d$), interactions between dislocations
can be neglected, and the energy per unit volume of
an array of low-angle grain boundaries is $f_{GB} = \fg L/(Al_b) = 2\fg/(l_d
\sqrt{3}l_b)$, where $\epsilon_d$ is the energy per unit length of a
dislocation.  Far from the boundary, $\theta = \phi_6$, and both $\theta$
and $\phi_6$ undergo the same jump $\Delta \theta$ across the boundary.
This jump was calculated in Ref. \onlinecite{KamNel}.
It is $b/l_d\sqrt{3}$ where $b = d$
is the magnitude of the Burgers vector.  The chiral energy of an array of
grain $N$ boundaries is thus
\begin{equation}
\Ha^* = N A {\gamma b\over l_d\sqrt{3}} = V{\gamma b\over \sqrt{3} l_b l_d}
,
\end{equation}
where $\gamma = \gamma_{\theta} + \gamma_{\phi}$ and $V = NA l_b$ is the
volume.
Thus, the energy per unit volume of the
moir\'{e} phase when dislocations interactions are ignored is
\begin{equation}
f_{m} = {1\over \sqrt{3}l_d l_b}(2 \fg - \gamma b ).
\end{equation}
This energy becomes negative, and the moir\'{e} state becomes energetically
preferable to the ordered phase when $\gamma >\gamma^m$, where
\begin{equation}
\gamma^m = 2 \fg/b .
\end{equation}
This result does not depend on the particular type of dislocation lattice
formed in the grain boundary.  For example, if the grain boundary consists
of identical orthogonal grids of dislocations with separation $l_d$ between
dislocations then $L= 2 l_d$, $A = l_d^2$, and $\Delta \theta = b/l_d$ so
that $f_m = (2 \fg - \gamma b )/(l_d l_b )$,
again producing $\gamma^m = 2 \fg/b$.
\par
The ordered phase becomes unstable with respect to the formation of the
moir\'{e} phase when $\gamma$ exceeds $\gamma_m$.  Our task, therefore, is
to calculate the energy per unit length of a screw dislocation $\fg$.
We follow closely the procedure of Ref. \onlinecite{Kos} appropriately
generalized to include $\theta$ as an independent variable.  We introduce
$w_{\gamma i}$, which is equal to $\partial_{\gamma} u_i$ away from
defects, where $i = 1,2$ and $\gamma = x,y,z$. We also introduce the
dislocation density,
\begin{equation}
\alpha_{\gamma i} (\xv ) = \sum_n
\int ds\, t_{n, \gamma} (s) b_{n,i} \delta^{(3)} ( \Rv_n ( s ) - \xv ) ,
\end{equation}
where $\Rv_n ( s )$ is the position vector of dislocation $n$  with
Burgers vector $\bv_n$ as a function of its arclength $s$ and $\tv_n (s) =
d \Rv_n (s)/ ds$ is its unit tangent vector.  The condition
$\oint_{\partial \Gamma} d u_i = b_i$ that the
integral of the changes in $\uv$ around a contour $\partial \Gamma$
enclosing a dislocation with Burgers vector $\bv$ be equal to $\bv$ then
implies the constraint,
\begin{equation}
\epsilon_{\mu\nu\gamma} \partial_{\nu} w_{\gamma i } = \alpha_{\mu i} (
\xv ) ,
\label{disconstr}
\end{equation}
on $w_{\gamma i}$.  To find the elastic energy associated with
dislocations, we need to minimize the energy $\Ha = \Ha_{\rm el} +
\Ha_{\theta}$ subject to this constraint.  Minimizing $\Ha$ with respect to
variations in $\theta$ and $\uv$, we obtain
\begin{eqnarray}
{\delta \Ha\over \delta
\theta}& = & -K^{\perp}_{\theta}\nabla^2_{\perp}\theta - K^{||}_{\theta}
\nabla^2_{||}\theta+g(\theta-\case{1}{2}\epsilon_{ij}w_{ij})=0 ,
\label{thetaeq}\\
{\delta \Ha\over \delta u_j }&=&
-\mu
(\partial_iw_{ij}+\partial_i
w_{ji})-\lambda \partial_j w_{ii}+K\partial_z^3w_{zj}\nonumber \\
& & + {g\over 2}(\epsilon_{ij}\partial_i\theta
-{1\over 2}\epsilon_{ij}\epsilon_{ij'}\partial_{i'}w_{i'j'}) = 0 .
\label{ueq}
\end{eqnarray}
After Fourier transforming, we can solve Eqs.\ (\ref{thetaeq}) and
(\ref{disconstr}) for $\theta$ and $w_{\gamma , i}$:
\begin{eqnarray}
\theta& = & {g\epsilon_{ij}w_{ij}\over 2(K_{\theta}^{||} q_{||}^2 +
K_{\theta}^{\perp} q_{\perp}^2 +g)}
\label{thetasolv}\\
w_{\gamma i}& = & {-i \epsilon_{\gamma\mu\nu}q_{\mu}\alpha_{\nu i}(\qv) \over
q^2}+iq_{
\gamma}\psi_i(\qv)
\label{wsolv}
\end{eqnarray}
where $\psi_i = i (q_i \sigma - \epsilon_{ij} q_j \pi )/ q_{\perp}^2$ defines
the longitudinal part, which is obtained by solving Eq.\ (\ref{ueq}).
Substituting Eq.\ (\ref{thetasolv}) into Eq.\ (\ref{ueq}) and using
$q_i \psi_i = i \sigma$ and $\epsilon_{li} q_l \psi_l = i \pi$, we obtain
\begin{eqnarray}
i\sigma&= & {1\over (2\mu +\lambda ) q_{\perp}^2
+Kq_z^4}\left[{\lambda\epsilon_{lm}q_{\perp}^2q_m
\alpha_{zl}(\qv)\over q^2}\right. \nonumber \\
& & \left. -{2\mu-Kq_z^2\over
q^2}\epsilon_{lm} q_lq_jq_z\alpha_{mj}\right] ,
\label{sigma}\\
i\pi&= & {1\over (\beta(q)+\mu) q_{\perp}^2+Kq_z^4}
\nonumber \\
& & \times
\left[-\beta(\qv )q_{\perp}^2(q_{\perp}^2q_{n}\alpha_{zn}(\qv)
+q_{\perp}^2q_z\alpha_{nn}(\qv) \right.  \label{pi} \\
& & +\left.{\mu+Kq_z^2\over q^2}q_zq_lq_i\epsilon_{lj}
\epsilon_{im}\alpha_{mj}-{\mu\over
q^2}q_lq_iq_{\rho}\epsilon_{lj}\epsilon_{j\rho \nu}
\alpha_{\nu i}(q)\right] \nonumber
\end{eqnarray}
where
\begin{equation}
\beta ( \qv ) = {1 \over 4}
{g (K_{\theta}^{||} q_{||}^2 + K_{\theta}^{\perp} q_{\perp}^2 )
\over g + K_{\theta}^{||} q_{||}^2 + K_{\theta}^{\perp} q_{\perp}^2 } .
\end{equation}
For a single screw dislocation aligned along the $x$-axis. $\alpha_{\nu i}
( \qv ) = 2 \pi b \delta_{\nu x} \delta_{ix} \delta ( q_x )$.
Using Eqs.\ (\ref{wsolv}, (\ref{sigma}), and (\ref{pi}), we obtain
\begin{eqnarray}
w_{yx}(q)& =&{-iq_z\alpha_{xx}(q)\over q^2}{Kq^2q_z^2\over
\mub ( \qv ) q_y^2+Kq_z^4} , \nonumber \\
w_{zx}(q)& = & {iq_y\alpha_{xx}(q)\over q^2}{\mub  ( \qv )q^2 \over
\mub( \qv ) q_y^2+Kq_z^4} ,
\label{w's}
\end{eqnarray}
where
\begin{equation}
\mub ( \qv )  = \mu + \beta ( \qv ) .
\label{mueff}
\end{equation}
Then, using Eq.\ (\ref{w's}) in $\Ha$, we obtain the energy per unit length
of a dislocation,
\begin{equation}
\fd = {1 \over 2} K b^2 \int {dq_y dq_z\over (2 \pi)^2} {q_z^2 \mub (\qv )
\over \mub ( \qv ) q_y^2 + K q_z^4} ,
\label{fd}
\end{equation}
where the integrals over $q_y$ and $q_z$ have respective upper cutoffs of
the inverse correlation lengths, $\xi_{\perp}^{-1}$ and $\xi_{||}^{-1}$.
This expression is identical to that obtained in Ref. \onlinecite{KamNel}
with $\mub ( \qv )$ replacing $\mu$.
\par
To evaluate $\fd$, it is convenient to express the integral in a unitless
form:
\begin{equation}
\fd = \epsilon_0 \int_0^1 dy \int_0^1 dz {z^2 ( 1 + f ) \over y^2 (1 + f )
+ \delta^4 z^4} ,
\end{equation}
where
\begin{equation}
\epsilon_0 = {K b^2 \over 8 \pi^2} {\xi_{\perp}\over \xi_{||}^3}
\end{equation}
and
\begin{equation}
f = {\beta_{||}^2 z^2 + \beta_{\perp}^2 y^2 \over 1 + \alpha_{||}^2 z^2 +
\alpha_{\perp}^2 y^2} .
\end{equation}
In the expressions for $\fd $ and $f$, we have introduced unitless ratios,
all of which are Ginzburg parameters measuring the ratio of a penetration
depth to a coherence length.  The parameter
\begin{equation}
\delta = {\sqrt{\lambda \xi_{\perp}} \over \xi_{||}}
\end{equation}
was introduced in Ref. \onlinecite{KamNel}.
The parameters
\begin{equation}
\alpha_{||} = {\lambda_{\theta}^{||} \over \xi_{||}} \qquad
\alpha_{\perp} = {\lambda_{\theta}^{\perp} \over \xi_{\perp}}
\end{equation}
and
\begin{eqnarray}
\beta_{||}& = & {1 \over 2 \xi_{||}}\sqrt{K_{\theta}^{||}\over \mu} =
{\lambda\over 2 \xi_{||}}\sqrt{K_{\theta}^{||}\over K} ,\nonumber \\
\beta_{\perp} & = & {1\over  2 \xi_{\perp}}
\sqrt{K_{\theta}^{\perp}\over \mu} = {\lambda\over 2
\xi_{\perp}}\sqrt{K_{\theta}^{\perp}\over K}
\end{eqnarray}
are Ginzburg parameters for twist penetration. Note that $\beta_{\perp}$
and $\beta_{||}$ are independent of $g$.
\par
Analytic evaluation of $\fd$ is difficult except for $g = 0$.  We will
content ourselves with the $\delta \rightarrow 0$ and $\delta \rightarrow
\infty$ limits for $g$ small and for $g - \infty$.  Since $d\fd/d g$ is
positive, $\fd$ will increase monotonically from the its value at $g=0$ to
its value at $g=\infty$.  As $g\rightarrow 0$, we find
\begin{enumerate}
\item $g \rightarrow 0$, $\delta = 0$
\begin{eqnarray}
\fd &= & {\pi \over 2} \epsilon_0 \delta^{-2} \left( 1 + {1 \over 8} { g
\over \mu} \right) \\
& = & {1 \over 16 \pi} {K^{1/2} \mu^{1/2} b^2 \over \xi_{||}}\left( 1 + {g
\over 8 \mu}\right) ,
\end{eqnarray}
and
\item $g \rightarrow 0$, $\delta = \infty$
\begin{eqnarray}
\fg& =&{\pi \over \sqrt{2}}\epsilon_0 \delta^{-3} \left( 1 + {3 \over 16} {g
\over \mu} \right) \nonumber \\
&=& {1 \over 8 \sqrt{2} \pi} {\mu^{3/4} K^{1/4} b^2 \over
\sqrt{\xi_{\perp}}} \left(1 + {3 \over 16} {g \over \mu} \right) .
\end{eqnarray}
\end{enumerate}
For $g = \infty$, we find
\begin{enumerate}
\item $g = \infty$, $\delta \rightarrow 0$
\begin{eqnarray}
\fd & = & {\pi \over 2} \epsilon_0 \delta^{-2} \nonumber \\
& & \times \left[{1 \over 2} \sqrt{1 +
\beta_{||}^2} + {1 \over 2 \beta_{||}} \ln (\beta_{||} +
\sqrt{1 + \beta_{||}^2} ) \right]
\end{eqnarray}
\item $g = \infty$, $\delta = \infty$
\begin{equation}
\fd = {\pi \over \sqrt{2}} \epsilon_0 \delta ^{-3} {1 \over 2} \int_0^1 dy
y^{-1/2} (1 + \beta_{\perp}^2 y^2 )^{3/4}
\end{equation}
\end{enumerate}
These results reduce to those of reference \onlinecite{KamNel} (when a
missing factor of $1/4$ is added).  $\fd$ increases smoothly
and monotonically with $g$. Its value at $g= \infty$ is finite and depends
on the Ginzburg parameters $\beta_{\perp}$ and $\beta_{||}$.  When
$\beta_{||}$ and $\beta_{\perp}$ are both zero, $f=0$, and $\fd(g)
= \fd(g = 0)$.  When these quantities are much greater than unit, $\fd(g =
\infty ) \gg \fd ( g = 0)$:
\begin{eqnarray}
{\fd ( g = \infty, \delta = 0)\over \fd ( g = 0 , \delta = 0 )} &
\rightarrow {1 \over 2} \beta_{||} , \qquad \beta_{||} \gg 1 \nonumber
\\
{\fd ( g = \infty, \delta = \infty)\over \fd ( g = 0 , \delta = \infty )} &
\rightarrow {1 \over 4} \beta_{\perp}^{3/2} , \qquad \beta_{\perp} \gg 1 .
\end{eqnarray}
Thus, large $g$ and large angle elastic constants $K_{\theta}^{||}$ and
$K_{\theta}^{\perp}$ lead to large values of $\fd$ and suppress the
formation of the moir\'{e} phase.
\section{Discussion and Review}
\par
In this paper, we have developed a model for chiral discotic columnar
liquid crystals, and we have investigated its instability toward the
formation of two types of structurally chiral phases: the soliton phase
and the moir\'{e} phase.  Chirality in our model, which restricts the
average molecular normal to be along the columnar axis,
gives rise to two kinds of chiral interactions, one tending to rotate
molecules within a column and the other tending to rotate the columns
themselves.  These two interactions are characterized by respective
coupling strengths $\gamma_{\theta}$ and $\gamma_{\phi}$.  There is an
energy cost associated with rotation of molecules relative to the lattice
characterized by a coupling constant $g$.  When $g$ is small, rotation of
molecules within columns with a fixed lattice structure is possible.  This
is the soliton phase that develops for molecular chiral coupling constant
$\gamma_{\theta}$ greater than a critical value $\gamma_{\theta}^s \sim
\sqrt{g}$.  When $g$ is larger formation of the moir\'{e} phase is favored
for $\gamma_{\theta} + \gamma_{\phi} > \gamma^m$ where $\gamma^m$ is a
smooth function of $g$ that is finite in the $g \rightarrow \infty$ limit.
\par
We have focussed on the instabilities toward two possible structurally
chiral phases.  A full phase diagram for the model and indeed for real
chiral discotic systems can be quite complex with mixed soliton-moir\'{e}
phases.  Additional phases can occur if the constraint that the molecular
normal (the Frank director) be parallel to the columnar axis be relaxed.
In particular, tilt-grain-boundary phases and smectic-$C^*$-like phase in
which the director tilts relative to the columnar axis and rotates in a
helical fashion along that axis can occur.  These more complex phase are
currently being investigated.
\par
We are extremely grateful for many conversations with  and continuous support
from Randall Kamien.  We also acknowledge useful converstions with Tim
Swager. This work primarily by the MRSEC Program of the National Science
Foundation under Award Number DMR96-32598.

\end{document}